\documentclass[12pt]{spieman}
\usepackage{newtxmath}
\usepackage{cleveref}
\DeclareRobustCommand{\VAN}[3]{#2}
\let\VANthebibliography\thebibliography
\def\thebibliography{\DeclareRobustCommand{\VAN}[3]{##3}\VANthebibliography}
\usepackage{soul}
\usepackage[dvipsnames]{xcolor}

\usepackage{svg}
\usepackage{graphicx}	
\usepackage{amsmath}	
\usepackage{xcolor}		
\usepackage{physics}	
\usepackage{multicol}   
 
\usepackage{nicefrac}
\usepackage[switch]{lineno}

\usepackage{csquotes}
\usepackage{url}
\usepackage{doi}
\usepackage{glossaries}
\usepackage{gensymb}
\usepackage{booktabs}
\usepackage{marvosym}

\newcommand\extrafootertext[1]{%
    \bgroup
    \renewcommand\thefootnote{\fnsymbol{footnote}}%
    \renewcommand\thempfootnote{\fnsymbol{mpfootnote}}%
    \footnotetext[0]{#1}%
    \egroup
}

\title{Mobile Intensity Interferometer for Stellar Observations (MI\textsuperscript{2}SO)}

\author[a,*]{Christopher Ingenh\"utt}
\author[a
]{Pedro Batista}
\author[a]{Gisela Anton}
\author[a]{Alison Mitchell}
\author[a]{Naomi Vogel}
\author[a]{Adrian Zink}
\author[a,b]{Andreas Zmija}
\author[a]{Stefan Funk}
\affil[a]{Erlangen Centre for Astroparticle Physics, Friedrich-Alexander-Universit\"at Erlangen-N\"urnberg, Nikolaus-Fiebiger-Str. 2, Erlangen 91058, Germany}
\affil[b]{Nice Institute of Physics, National Center for Scientific Research, 17 rue Julien Lauprêtre, 06200 Nice, France}

\begin{document}
\label{firstpage}
\maketitle

\begin{abstract}
In recent years, intensity interferometry has seen renewed interest and successful application at Imaging Atmospheric Cherenkov Telescope arrays. These measurements are usually performed during bright moon periods while the instruments' primary purpose---gamma-ray observations---cannot be fulfilled. The Mobile Intensity Interferometer for Stellar Observations was designed as a proof of concept for a purpose-built intensity interferometer.
Using acrylic Fresnel lenses $1\,$ m in diameter with $1.2\,$ m focal length, a compact, economical and lightweight design was realised. The detector fixture allows for translation in the z-axis to adjust for measurements at different wavelengths (and therefore focal points) and easy swapping of the detector in its entirety. 
Both mobility and scalability in quantity of this design allow for specific targeting of projected baselines and orientations based on the target. Particularly for potential binary systems, selective coverage of a target's u-v plane is essential to probing the characteristics accurately.
A first campaign demonstrated the capability of these Fresnel lens telescopes by measuring the spatial coherence curve of Arcturus ($\alpha$ Boo). In an observation time of less than $11\,$ h, the angular diameter was measured with milliarcsecond precision, in agreement with the values in the literature.
\end{abstract}
   

\extrafootertext{\noindent{\textbf{*}Direct all correspondence to: Christopher Ingenh\"utt,  \linkable{christopher.ingenhuett@fau.de} }}


\section{Introduction}
Since its conception by Hanbury Brown and Twiss in 1956\cite{hanbury_brown_test_1956}, stellar intensity interferometry (SII) has proved to be a great technique for measuring the angular diameter of stars.
Unlike amplitude interferometry, it is insensitive to atmosphere fluctuations while making astronomical observations at optical wavelengths. However, intensity interferometry is limited by the detector collection area and sampling rates. 

After multiple decades of slow development, the technique is experiencing renewed interest with the advent of Imaging Atmospheric Cherenkov Telescopes (IACTs). Arrays of these telescopes are used for gamma-ray observations of the Cherenkov light of particle showers caused by gamma rays interacting in the atmosphere. This technique is challenging to use with strong night-sky background brightness, which is why observations are usually suspended on nights with a (nearly) full moon in the sky. This pause is now being used by all currently operating IACT arrays for intensity interferometry observations, see the Major Atmospheric Gamma Imaging Cherenkov Telescopes (MAGIC)\cite{abe_performance_2024}, Very Energetic Radiation Imaging Telescope Array System (VERITAS)\cite{abeysekara_demonstration_2020}, and High Energy Stereoscopic System (H.E.S.S.)\cite{vogel_simultaneous_2025,zmija_first_2024}.  
Hanbury Brown himself noted in his book \cite{hanbury_brown_intensity_1974} that the telescope collection area would be one of the first factors to improve the sensitivity of intensity interferometry. With collection areas on the order of $\sim\,$O($100\,$m\textsuperscript{2}), arrays of IACTs are well-suited to advancing the technique in this way.
However, SII should not necessarily depend only on operating IACTs, but in addition take a more direct approach with purpose-built SII detectors. This way, observation procedures, instrument design, and observation schedules can be optimized specifically for SII science goals.

\begin{figure}[!h]
    \centering
    \includegraphics[width=0.9\columnwidth]{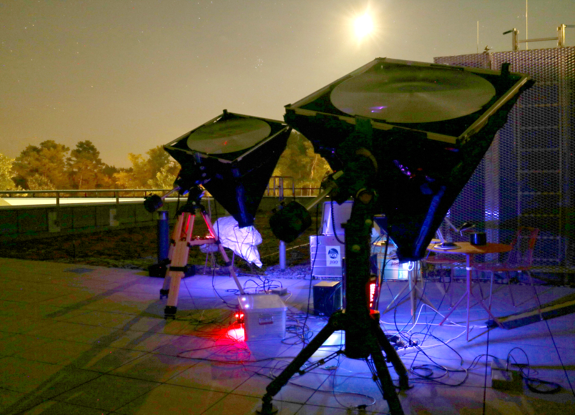}
    \caption{The MI\textsuperscript{2}SO telescopes in operation on the roof of the ECAP building during a night with bright moonlight.}
    \label{fig:miisopic}
\end{figure}

The Mobile Intensity Interferometer for Stellar Observations (MI\textsuperscript{2}SO) was developed at the Erlangen Centre for Astroparticle Physics (ECAP), in Germany, and present a new, cost-efficient alternative for stellar intensity interferometry measurements.
Figure \ref{fig:miisopic} shows the MI\textsuperscript{2}SO telescopes in operation during one of the observation nights.

Due to the inverse nature of the relation between physical size of a structure and required projected baseline to resolve it, large-collection-area telescopes are only sensitive to resolving certain aspects of a star system.
The projected baseline $b$ is the distance between the detectors in the projected viewing plane from the star.

\begin{figure}[!h]
\centering
    \includegraphics[width=0.9\columnwidth]{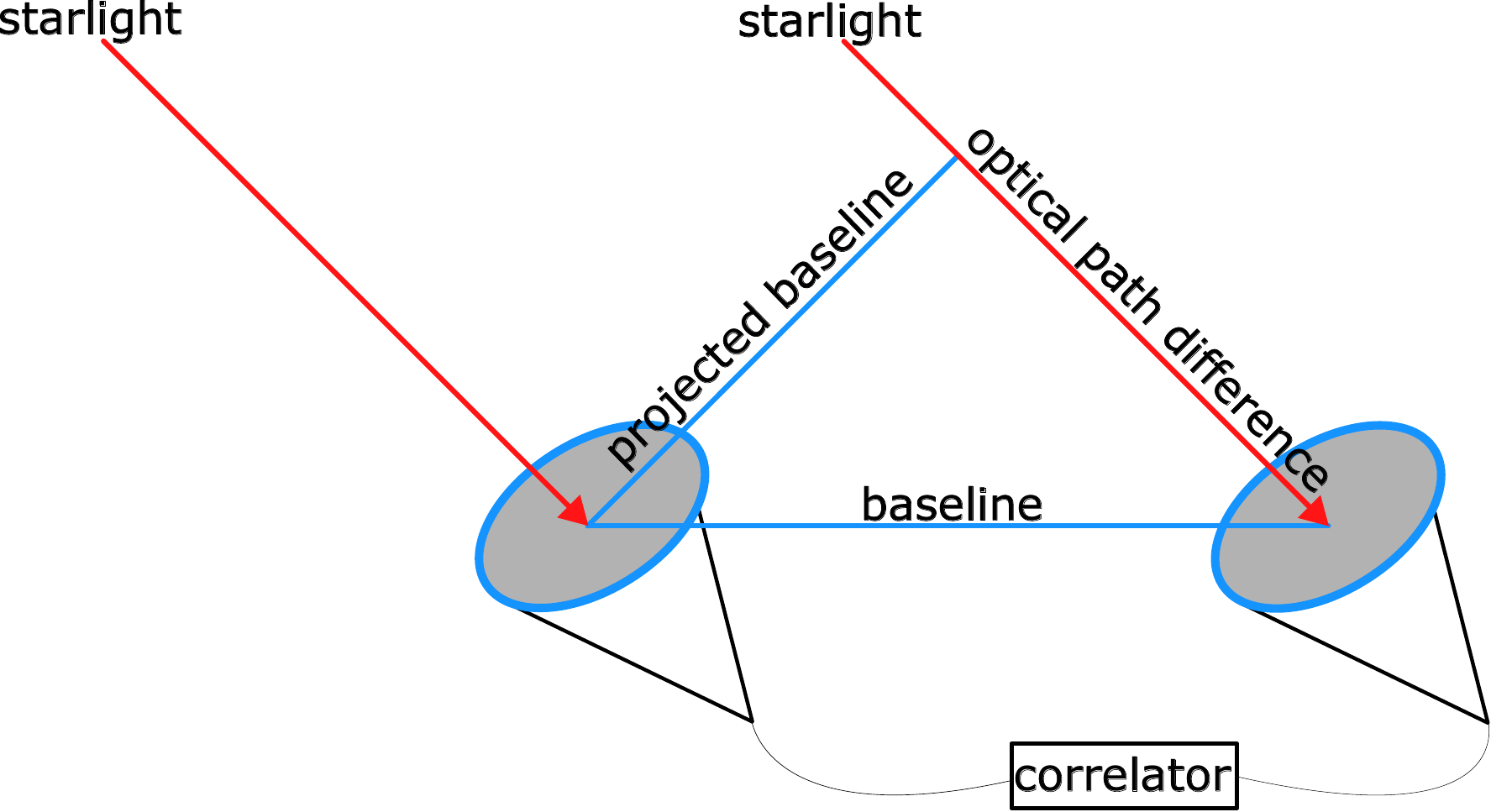}
    \caption{Projected baseline and optical path delay illustrated for off-zenith observations. Figure adapted from \citenum{noauthor_basics_nodate}.}
    \label{fig:geometry}
\end{figure}

This projected baseline is determined by using the pointing direction of the telescopes towards the star's position in the sky and their physical distance from each other. Figure \ref{fig:geometry} shows the relationship between the physical distance between the telescopes (often called baseline), projected baseline and optical path difference (which introduces a delay in coincident photon arrival times in one telescope relative to the other). 
As such, the limits for the values of $b$ are the aperture diameter of the telescope and their physical distance. The lower limit is reached in observations close to the horizon with the star's azimuth aligned with the connecting line of the telescopes, whereas the upper limit is reached for observations at zenith or with the star's azimuth perpendicular to the connecting line.

A set of stationary telescopes therefore needs high multiplicity to achieve dense coverage of the projected viewing plane, or u-v plane. The density and also range of coverage is especially relevant for targets with characteristics at different scales, for example binary systems with the individual stars' diameters and their separation. In addition, a large collection area of each telescope results in small-scale variations in spatial coherence being washed out as intensity is integrated over the entire collector. One solution to recover sensitivity for small-scale variations is masking parts of the IACT dish as multiple individual collectors (see \citenum{delgado_intensity_2021}), but this is not applicable to all current IACT designs as it relies on active mirror control aligning individual mirror facets with individual pixels of the Cherenkov camera.

Another solution is adding smaller telescopes to the array, as the lower baseline uncertainty resulting from a smaller collector area allows precise measurements of variations across projected baselines shorter than the IACTs' diameter. Multiplicity and mobility of such smaller telescopes (enabled by low cost and weight) would in turn allow for dense and arbitrary coverage of the u-v plane, effectively extending the dynamic range of resolution.

Achieving high-significance results with SII will always be limited by sensitivity, and hence by observation time. In this regard, dedicated instruments are preferable over "borrowing time" on instruments with other purposes (e.g. IACTs during bright moon nights), provided the construction and operation costs of these dedicated intensity interferometers is low or at least comparable to the operating costs of the respective borrowed instruments in SII mode.

MI\textsuperscript{2}SO was developed specifically to evaluate the feasibility of performing SII measurements with lightweight, cost-effective telescopes to allow both quantity scalability and mobility to optimize coverage of specific baselines of interest for any given target. Furthermore, the limitations and potential applications of intensity interferometers using relatively small Fresnel lenses to augment measurements with large collection area telescopes, such as IACTs, are investigated.

Section \ref{sec:observables} will give a brief theoretical review of the physical concepts necessary for intensity interferometry.
The material and design details of the MI\textsuperscript{2}SO telescopes will be discussed in Section \ref{sec:meas_setup}. We detail the measurements and operations in Section \ref{sec:meas_prod}. The analysis of the data is described in Section \ref{sec:analysis} and the results of this are presented in Section \ref{sec:results}. Finally, a conclusion and evaluation of MI\textsuperscript{2}SO's potential is given in section \ref{sec:conclusion}.

\section{Theory and Measurement Setup}
\subsection{Theoretical Review}\label{sec:observables}
In this section we provide a short review of SII theory (see \citenum{hanbury_brown_intensity_1974} for further information).

The photon intensity at two detectors $I_1$ and $I_2$, will have a time-averaged second-order correlation given by

\begin{equation}\label{eq:g2_function}
    g^{(2)}\left( \boldsymbol{b}, \tau \right) = \frac{\langle I_1 \left(\boldsymbol{r}, t\right)  I_2 \left(\boldsymbol{r} + \boldsymbol{b}, t + \tau \right)\rangle}{\langle I_1 \rangle \langle I_2 \rangle},
\end{equation}

where $\boldsymbol{b}$ is the projected baseline, $\boldsymbol{r}$ the position of the detector, and $\tau$ is the time difference between the detected photon signals.
Subsequently, the second-order correlation $g^{(2)}\left( \boldsymbol{b}, \tau \right)$ can be related to the first-order correlation $g^{(1)}\left( \boldsymbol{b}, \tau \right)$ via the Siegert relation \cite{ferreira_connecting_2020, siegert_fluctuations_1943} given by

\begin{equation}\label{eq:siegert_relation}
    g^{(2)}\left( \boldsymbol{b}, \tau \right) = 1 + |g^{(1)}\left( \boldsymbol{b}, \tau \right)|^2.
\end{equation}

According to the Wiener–Khinchin theorem \cite{mandel_optical_1995}, the temporal dependency of the first-order correlation, $g^{(1)}\left( \tau \right)$, can be described by the Fourier transform of the frequency power spectrum.
Furthermore, within a coherence time $\tau_\mathrm{c} \approx \frac{\lambda^2_0}{c \Delta \lambda}$, a photon-bunching peak in $g^{(2)}(\tau)$ is expected, where $\lambda_0$ is the central wavelength during observations, $\Delta \lambda$ is the optical bandwidth, and $c$ is the speed of light.
A more general approach to define the coherence time $\tau_\mathrm{c}$ for an arbitrary wavelength spectrum is given as

\begin{equation}\label{eq:coh_time_integral}
    \tau_\mathrm{c,measured} \coloneq \int^{+\infty}_{-\infty} \left(g^{(2)}(\tau) - 1\right)d\tau .
\end{equation}

This definition also addresses the issue of the physical coherence time typically being orders of magnitude smaller than the time resolution of the detectors.
The spatial dependence $g^{(1)}\left( \boldsymbol{r} \right)$ can also be described by a Fourier transform of the spatial intensity distribution of the source, according to the van Cittert-Zernike theorem \cite{mandel_optical_1995}.
\begin{equation}
    \label{eq:vcztheorem}
    g^{(1)}(r_1,r_2,0)=\text{e}^{ik(r_2-r_1)}\frac{\int_\sigma I(r^\prime)\text{e}^{-ik(s_2-s_1)\cdot r^\prime}d^2r^\prime}{\int_\sigma I(r^\prime)d^2r^\prime}
\end{equation}

If the spatial intensity of the source is expected to be distributed in a uniform disc of angular diameter $\theta$, the spatial first-order correlation at a projected baseline $b$ yields

\begin{equation}\label{eq:ud_correlation}
    g^{(1)}\left( b \right) = 2 \frac{J_1(\pi b \theta / \lambda)}{\pi b \theta / \lambda},
\end{equation}

where $J_1$ is the Bessel function of the first kind as a function of wavelength $\lambda$. 
Thus, by using two synchronized detectors with a linear response to light intensity, the angular diameter of a thermal light source (i.e. a star) can be probed:
\begin{equation}\label{eq:measg2}
    g^{(2)}\left(b,\tau=0\right) = \frac{\langle R_1R_2\rangle}{\langle R_1\rangle \langle R_2\rangle} = 1 + \left|g^{(1)}\left(b\right)\right|^2
\end{equation}
using the photon rates in the two detectors $R_1$ and $R_2$ in the respective time averages, as well as $g^{(1)}\left(b\right)$ from equation \ref{eq:ud_correlation}.

\subsection{The MI\textsuperscript{2}SO System}\label{sec:meas_setup}

\begin{figure}[!ht]
    \centering
    \includegraphics[width=0.9\columnwidth]{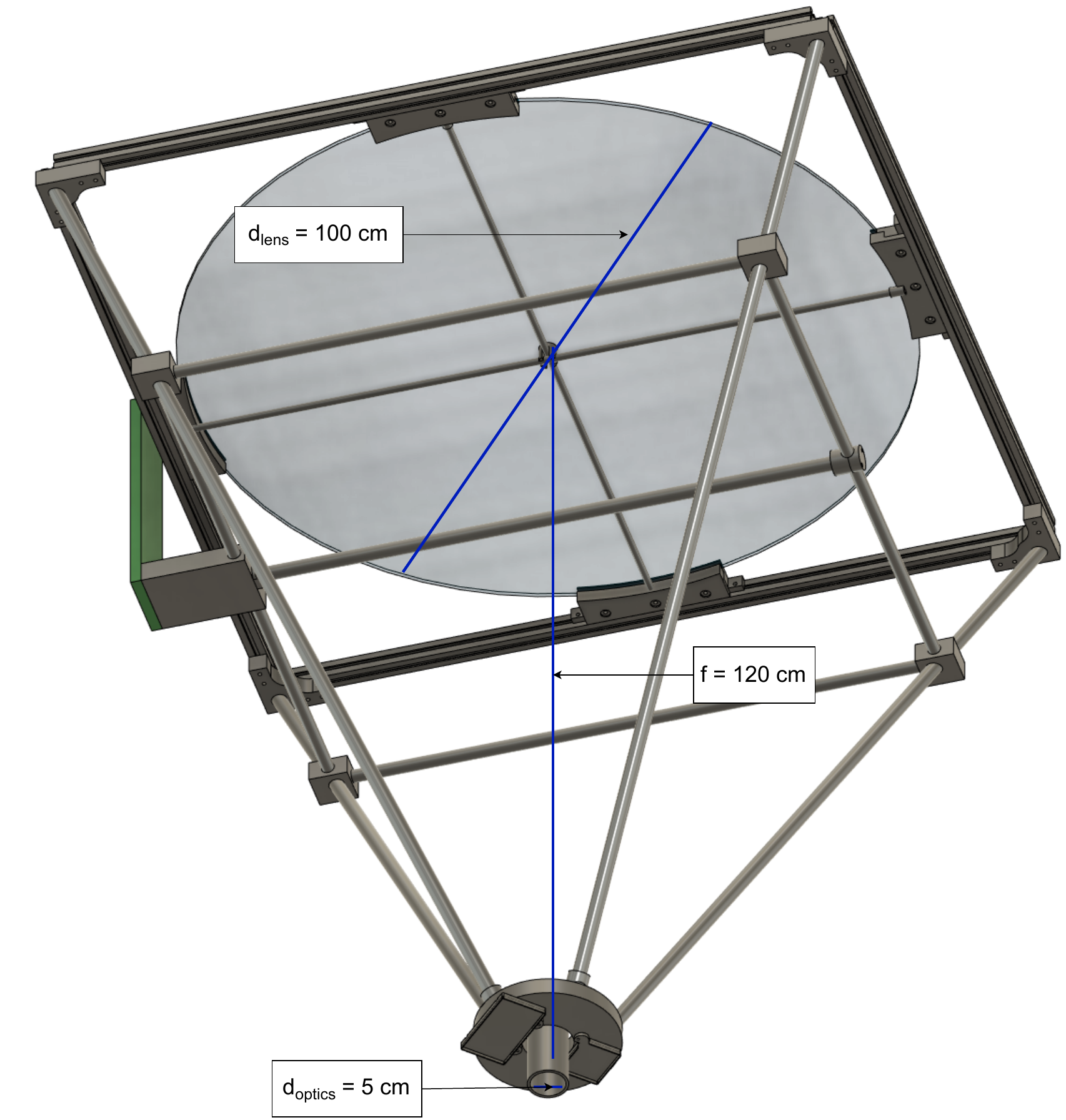}
    \caption{Schematic of a MI\textsuperscript{2}SO telescope facing up with dimensions of primary components indicated. The main lens holding frame is made of aluminium rails with 3D-printed PLA plastic two-part lens holders at the center of the rails and 3D-printed PLA plastic corner sockets into which $18\,$mm diameter carbon fiber rods are glued.}
    \label{fig:miiso-schem}
\end{figure}

In Figure \ref{fig:miiso-schem}, we present a schematic of a MI\textsuperscript{2}SO telescope and its dimensions. The system consists of two telescopes, each with a $1\,$m diameter, $1.2\,$m focal length Fresnel lens, held by a standard aluminium profile frame of 1\,m x 1\,m. 
The telescope structure is made of carbon fiber rods and is covered with dark cloth to prevent stray light from reaching the detector. 

A secondary stabilizing frame made of carbon fiber rods and plastic components increases the stability of the telescope.
 The optical detector is placed at the focal point supported by a plastic fixture that allows manual adjustments of its position for focussing.
The 90\,\% containment diameter $d_\mathrm{90}$ of the Fresnel lenses was measured in the laboratory to a size of $d_\mathrm{90} = 8.06 \pm 0.24$\,mm \cite{ingenhutt_intensity_2025}.
Therefore, a $9\,$mm diameter pinhole is placed in front of each PMT to greatly decrease contamination of the data with night sky background (NSB) noise at an insignificant loss of signal, assuming precise tracking of the target star. 

Because we do not collimate or parallelize the lightbeam after focussing through the primary Fresnel lens, we use wide-band optical filters with a central wavelength of $\lambda_0 = 655\,$nm and a width of $\Delta\lambda = 47\,$nm. Such optical filters are less affected by divergence of the angle of infalling light, and the loss in coherence is compensated by the gain in photon rate. These filters are placed between the PMT and pinhole and can be swapped depending on the target's emission spectrum.

The two telescopes are mounted on a Skywatcher EQ8-R (manual \citenum{sky-watcher_skywatcher_nodate-1})
and Skywatcher EQ8 (manual \citenum{sky-watcher_skywatcher_nodate}) 
mounts, respectively.
For star tracking, the mounts receive input from guiding cameras mounted into the corner of the lenses' frames. The guiding cameras are a Starlight Xpress Lodestar Pro (manual \citenum{xpress_starlight_2022}) and a Starlight Xpress Lodestar X2 (manual \citenum{xpress_starlight_2014}). The choice of mounts and guiding cameras was primarily driven by local availability. 
No visible impact of tracking inaccuracies on photon rates could be observed during observation nights. Even in the case of short-term loss of tracking due to clouds obscuring the star, photon rates recovered to the previous level as tracking was restored.

The chosen optical detectors are two Hamamatsu H15461 photomultiplier tubes (PMTs). These PMTs have a detection efficiency of $\sim\!40\,\%$ at wavelengths around $\lambda_0$ according to their datasheet \cite{kk_photomultiplier_2023}. 
Since SII measurements can be heavily limited by the time resolution of the system, the transit time-spread (TTS) of our optical detectors and detection electronics was measured to be $\textrm{TTS}_{\textrm{FWHM}} = 0.75\,$ns at full-width half-maximum \cite{blass_charakterisierung_2024}, providing sufficient time resolution for SII and comparable uncertainties as those resulting from the sampling speed ($1.25\,$GHz resulting in timebins of $0.8\,$ns) or time dispersion of the lens (plane lens with $1\,$m diameter $1.2\,$m focal length giving a maximum dispersion of $346\,$ps).

To the total weight of $\sim\!11\,$kg, the lens and its aluminium frame contribute $\sim\!3\,$kg each, the carbon fiber rods $\sim\!1.5\,$kg, the various plastic connectors and holders $\sim\!2\,$kg. The rest of the contributions depended on mounts chosen (for the mounting rail required) and detectors used and potential extra components necessary close to the detectors. For the PMTs used, a voltage control box and signal amplifier were mounted to the back of the detector fixture to minimize cable lengths between the PMTs and these extra components. Holders for these can be seen in the schematic (figure \ref{fig:miiso-schem}) at the bottom.

\subsection{Measurement Procedure}\label{sec:meas_prod}
Intensity interferometry measurements were conducted over six nights in July and September 2024, taking place on the roof of the Erlangen Centre for Astroparticle Physics (ECAP) located at $49\degree34'53.0''$N $11\degree01'50.3''$E.

For the selection of targets, a list of stars with magnitude $B \lesssim 2$ was created, possible observing hours were calculated for nights after nautical darkness (Sun altitude $\leq -12 \degree$), and star altitudes $\geq 20\degree$. 
Furthermore, stars with higher emissions at longer wavelengths ($\sim\!650$\,nm) are preferred, due to the choice of filters used in the MI$^2$SO telescopes (see Section \ref{sec:meas_setup}).

To guide target selection, an estimate of the expected significance of a measurement of $g^{(2)}$ $n$ was made at a theoretical baseline $b = 0$, after observing for a period of time $T$. This significance $n$ can be estimated following the expression:

\begin{equation}\label{eq:significance}
    n = \sqrt{
            \frac{ \lambda_0^4 R_\star^2 T }{ 4\sigma_t c^2 \left( \Delta\lambda \right)^2 \left(1 + \frac{R_\mathrm{NSB}}{R_\star} \right)}
            },
\end{equation}

where $R_\star$ and $R_\mathrm{NSB}$ are the star and night-sky background (NSB) rates estimated at $\lambda_0$ and $\Delta\lambda$, and $\sigma_\mathrm{t}$ is the time resolution of the detector.

\begin{figure}[!h]
    \centering
    \includegraphics[width=0.9\columnwidth]{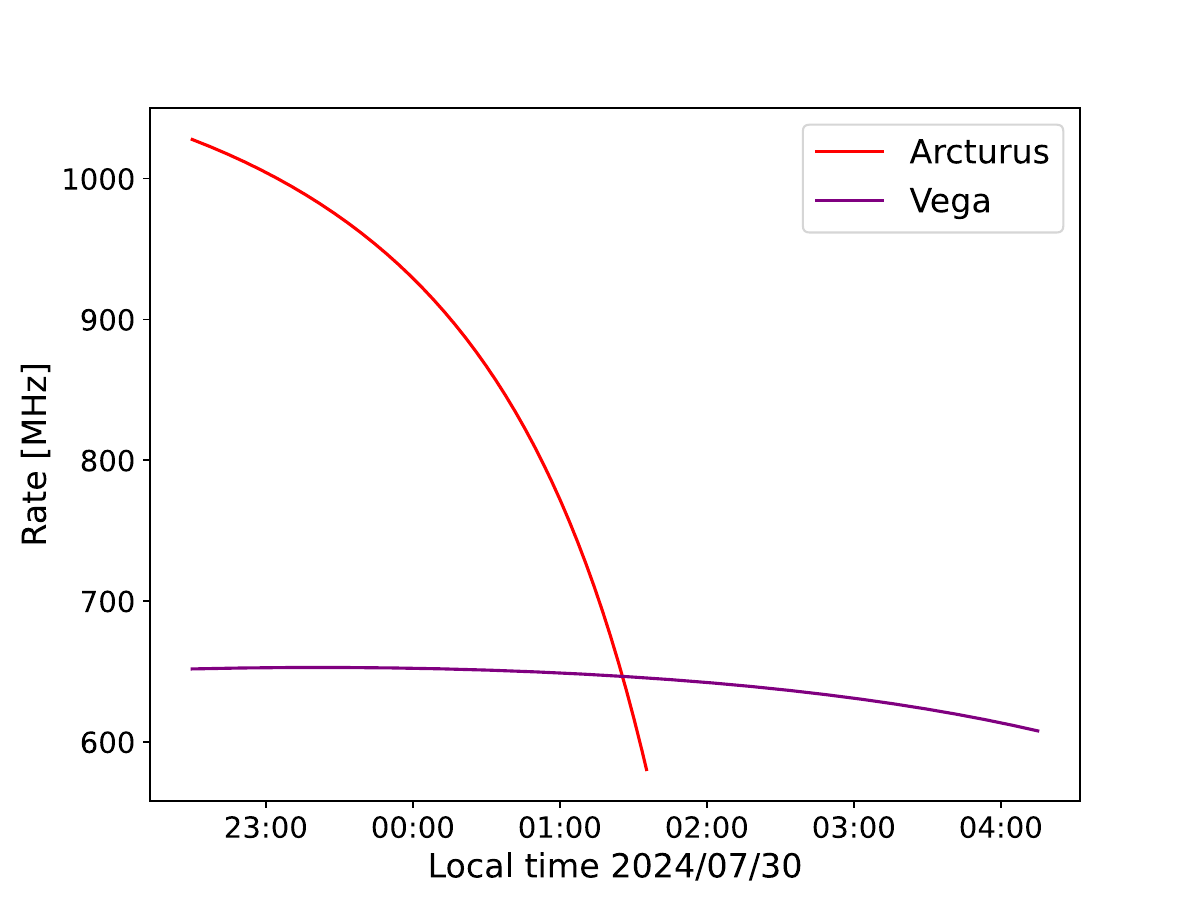}
    \caption{Predicted photon rates for Arcturus and Vega in the observation night of July 30th 2024. Colors of the curves were chosen to emphasize effective temperature and therefore peaking wavelength difference between Arcturus ($4286\,$K and $676\,$nm) and Vega ($9650\,$K and $300\,$nm). The simulation uses manufacturer values for lens transmission, filter transmission and PMT quantum efficiency as well as literature values for altitude and wavelength dependent atmospheric absorption, see \citenum{stubbs_toward_2007}.}
    \label{fig:ratecomp}
\end{figure}

After estimating expected photon rates of the twenty brightest stars in the system using the wavelength-dependent optical filter's transmission profile, lens transmission and PMT detection efficiency, as well as atmospheric absorption at each potential target's altitude during the observation period, the stars Arcturus \cite{ramirez_fundamental_2011} and Vega \cite{monnier_resolving_2012} were selected as the best candidates for observation. Figure \ref{fig:ratecomp} shows a comparison between Arcturus's and Vega's predicted photon rates for the night of July 30th, 2024.
The time-dependent rates are estimated by taking into account the system's efficiency as well as expected atmospheric absorption at the calculated altitudes of the targets.

\subsubsection{Relative positions of the telescopes}\label{subsec:tel_position}

The relative position of the telescopes was optimized to minimize baseline variation and to select distinctly different points on the expected spatial coherence curve. This is primarily an attempt to compensate for the low signal-to-noise figure necessitating averaging the $g^{(2)}$ function over large numbers of files, expected to be ideally a full night's worth for one datapoint. As such, Earth's rotation is impractical to use to create different projected baselines and the telescopes' relative positions were adjusted every two observation nights. 

\begin{figure}[!h]
    \centering
    \includegraphics[width=0.9\columnwidth]{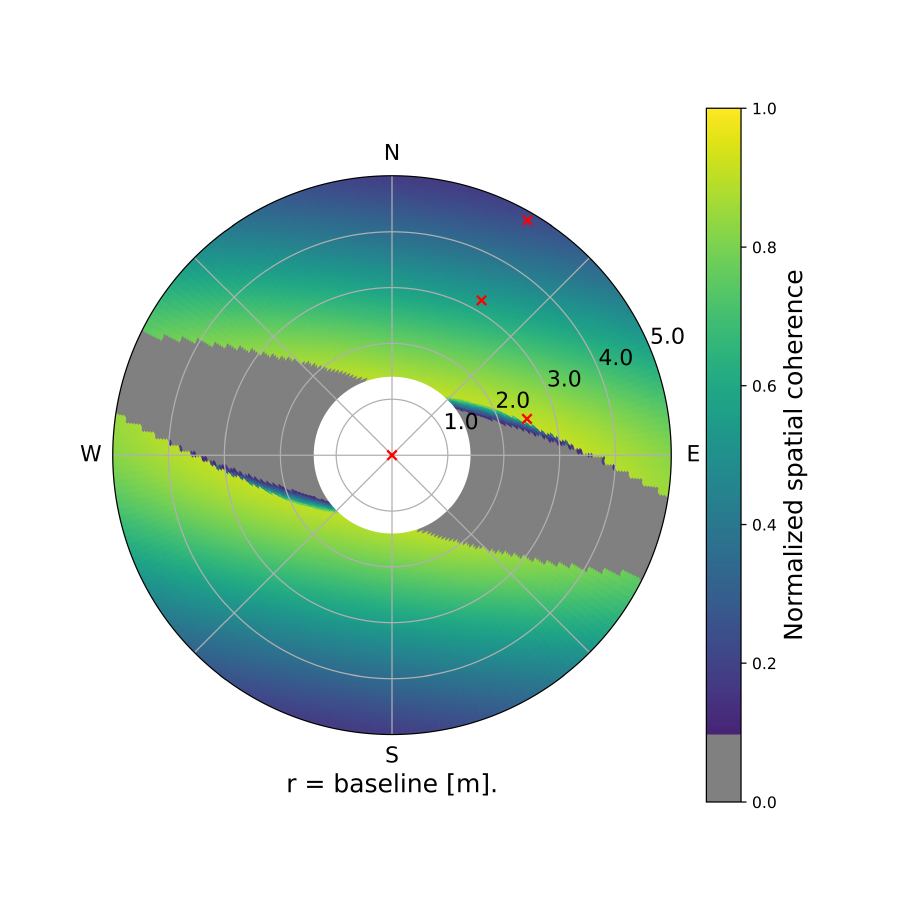}
    \caption{Average spatial coherence (color scale) expected for the target star Arcturus as a function of the position of the second telescope, over the course of the observation night of July 30th, 2024, with the first telescope in the centre. To exclude telescope orientations in which one telescope casts a shadow on the other, spatial coherence is set to zero for times during the night when the projected baseline is smaller than the telescope frame diameter. The grey area in the plot highlights orientations containing some fraction of observation time affected by this shadow. The telescope positions that were used are indicated in red.}
    \label{fig:compassplot}
\end{figure}

In Figure \ref{fig:compassplot} the average spatial coherence values are shown for different telescope configurations. The telescope positions used are indicated in red. One telescope is always in the center, the second telescope was placed at three different positions. Two nights of observations were carried out at each configuration to probe different parts of the spatial coherence curve (relative positions specified in table \ref{tab:obs_times}). 
In order to calculate the average spatial coherence of Arcturus, the projected baseline is simulated based on the trajectory over one night. Thus, the figure only shows the expected measured values if each relative telescope orientation was realized on the same night and is not meant to be accurate for all of them.

\begin{figure}[!h]
    \centering
    \includegraphics[width=0.9\columnwidth]{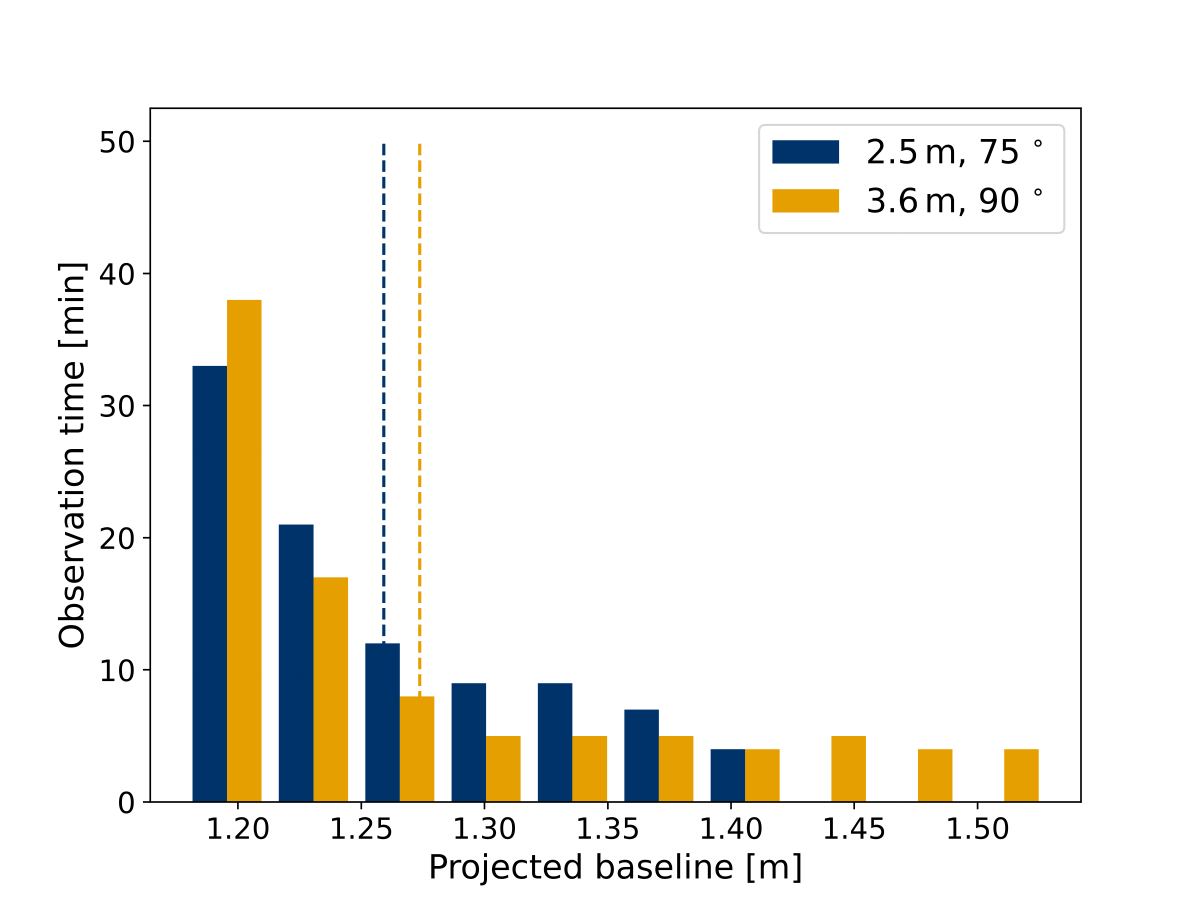}
    \caption{Histogram of observation time in minutes at different projected baselines for two potential orientations of the telescope pair (described by distance and meridian angle). While the average projected baseline, indicated by dashed lines, is very similar between the two orientations, the variation in baseline is significantly larger in the $3.6\,$m $90\,\degree$ orientation in orange.}
    \label{fig:telorient}
\end{figure}

As the previously explained compass plot simulation only shows the average normalized spatial coherence over the course of the night, it is primarily used to pick distinctly different projected baselines. A secondary simulation is used to determine the telescope orientation with the lowest baseline variation for a given aim in average spatial coherence. Figure \ref{fig:telorient} shows the projected baseline variation for two different orientations simulated for the observation night of September 19, 2024 as a histogram of observation time in minutes at different projected baselines in simulations. The mean projected baselines are indicated by dashed lines. The configuration at $3.6\,$m distance and a meridian angle of $90\,\degree$ (orange in the histogram) would have one telescope in the middle of figure \ref{fig:compassplot} and the second telescope on the W-E gridline at $3.6$. The position of the second telescope in the configuration at a distance of $2.5\,$m and meridian angle of $75\,\degree$ (blue in the histogram) is marked with a red x between the 2.0 and 3.0 gridlines. Although the average spatial coherence values would be very similar between the two configurations, the variation in the projected baseline differs as can be seen in figure \ref{fig:telorient}. The configuration with lower variation is chosen for the September measurements.

\begin{table}[!h]
    \centering
    \begin{tabular}{ccccc}
        \toprule
           & Arcturus (h) & Vega (h) & distance [m] & meridian angle [deg] \\
        \midrule
         2024-07-18 & 2.48 & 0.92 & 3.2 & 30 \\
         2024-07-28 & 1.19 & --- & 3.2 & 30\\
         2024-07-29 & 2.13 & 1.70 & 4.85 & 30\\
         2024-07-30 & 2.86 & 1.81 & 4.85 & 30\\
         2024-09-19 & 0.97 & --- & 2.5 & 75\\
         2024-09-21 & 1.21 & --- & 2.5 & 75\\
         \midrule
         \textbf{Total} & \textbf{10.84} & \textbf{4.43} \\
         \bottomrule
    \end{tabular}
    \caption{Total amount of photo-current data per night and target in hours. Distance between the telescopes and meridian angle are also listed.}
    \label{tab:obs_times}
\end{table}

Finally, Table \ref{tab:obs_times} shows the total number of observing hours per night, per target. Relative telescope orientation is indicated in the last column.
Arcturus was the main target for observations, with a total of 10.84 hours of observation, because of its higher expected emission at the filters' wavelength. 
Vega was observed on nights after Arcturus had set and was not observed on some nights due to adverse weather conditions, giving a total of 4.43 hours of observation.
For the nights of 18th and 28th of July, the telescopes were at a distance of 3.2\,m at meridian angle of $30\degree$ NNE, and at a distance of 4.85\,m for the nights of 29th and 30th in the same orientation.
On the 19th and 21st of September, the telescopes were at a distance of 2.5\,m at meridian angle of $75\degree$ ENE.

\section{Data Analysis and Results}
\subsection{Data Analysis}\label{sec:analysis}
The photon currents from the two synchronized detectors are amplified and then digitized using a high-speed digitizer card (Spectrum M4i 2211-x8 with a sampling rate of $1.25\,$GHz and an 8-bit resolution \cite{spectrum_instrumentation_m4i22_2025}), storing the data as files containing 2 Gigasamples per channel. This filesize was chosen to make full use of the digitizer card's internal buffer ($4\,$GB). To mitigate correlated electronic noise---specifically at zero delay---the signal cables to the two detectors have different lengths, shifting the photon bunching peak away from the correlation center.

Each PMT current file is timestamped, and the mean ADC count per channel is recorded. After an offset correction, the mean ADC counts are used as an estimate of photon rates over $\sim\!1.7$\,s acquisition windows. The cross correlations $G_{12}^{(2)}$ (see Eq. \ref{eq:measg2}) between channels are calculated using an FFT-based cross-correlation method for every file at a correlation length of $5 \times 10^{4}$ samples (40\,$\upmu$s) for computational efficiency.

The baseline dependent second-order correlation $g^{(2)}(b)$ is obtained as the normalized sum of weighted individual cross-correlations between detectors 1 and 2 $G_{12}^{(2)}$.
As weights, we use the inverse of the standard deviation of the cross-correlation between the corresponding 1.7\,s waveforms from each detector.
We exclude the zero-delay section of cross-correlation files from the evaluation of the standard deviation due to large amounts of correlated electronic noise.
This weighting aims to evaluate the contribution of each  $\sim1.7$\,s window to the full measurement's summed correlation function.
Optical path differences between telescopes are corrected by phase-shifting (rolling) each file by the nearest integer of sampling bins corresponding to the path-length difference. 

Since the physical distance and orientation of the two MI\textsuperscript{2}SO telescopes were constant for two of the six nights each, data from nights with the same setup can be combined to improve statistics, provided that the nights are close enough in time that the target's path in the sky leaves their projected baseline relatively constant between them. 
Data for the nights of July 18 and 28 were \textit{ not} combined due to a large gap of unfavourable weather between them. 
Data from the nights of July 29th and 30th are combined to form the "long baseline" dataset, data from the nights of September 19th and 21st form the "September" dataset. 
To combine the $g^{(2)}(b)$ of different nights, the mean photon rate product times the observation time ($R_1\cdot R_2\cdot N_\textrm{files}\cdot 2^{31}\cdot 0.8\cdot10^{-9}$\,s) was used as a weight for each night before normalization. This weight represents the dependence of the SNR as outlined in Eq. \ref{eq:significance}.

\begin{figure}[!h]
    \centering
    \includegraphics[width=0.9\columnwidth]{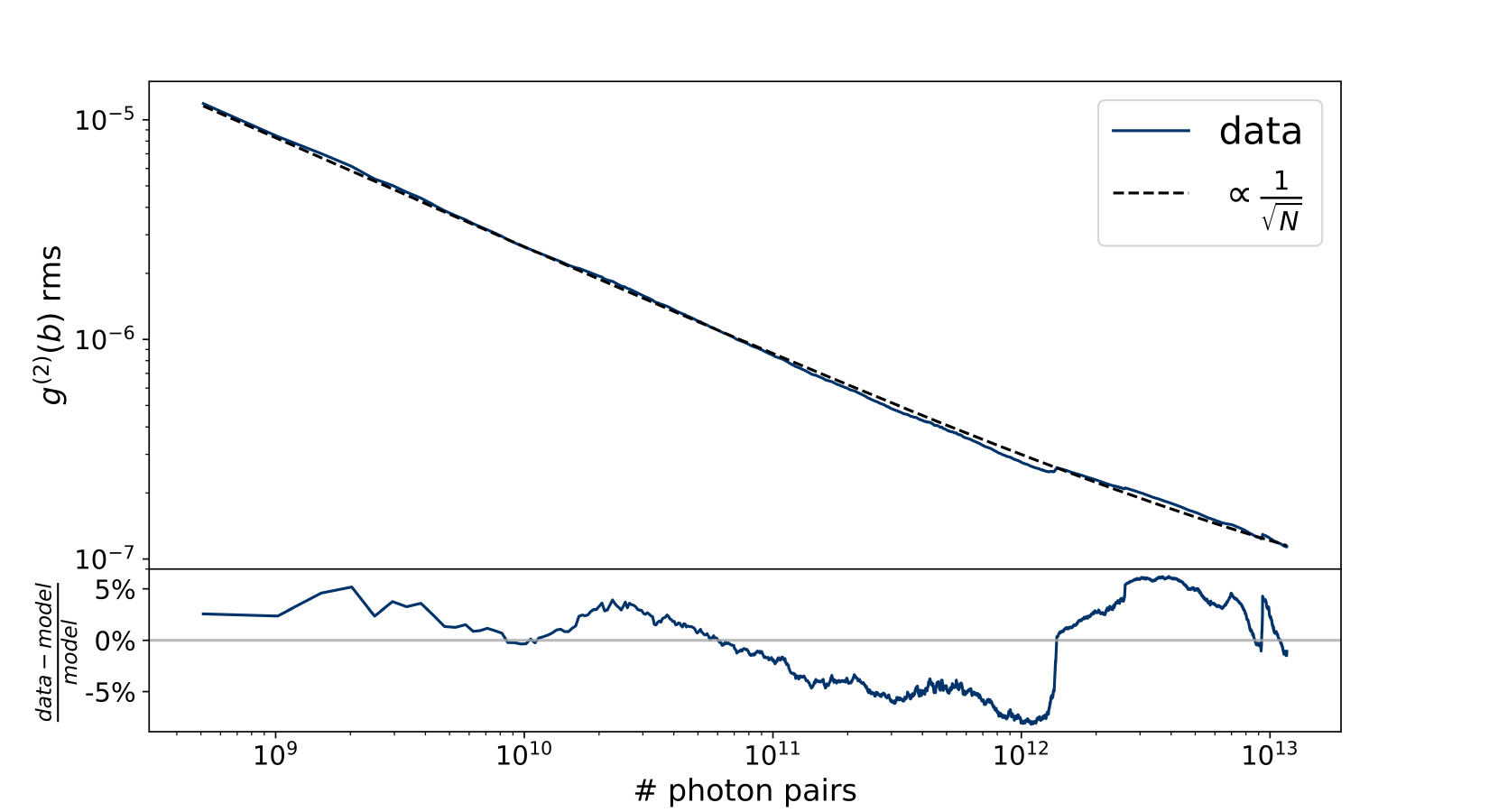}
    \caption{Top: Cumulative error on $g^{(2)}$ over the number of photon pairs for all Arcturus data. Central area of the $g^{(2)}$ containing correlated internal digitizer card noise and the photon bunching peak is excluded to reflect only the incoherent photon noise. Expectation fit is based on Poissonian noise assumption. The fitted function is $\propto1/\sqrt{N}$. Bottom: Residual estimation between noise data and the aforementioned function. Discontinuities occur at each night's first file being added to the cumulative analysis.}
    \label{fig:cumrmsplot}
\end{figure}

Several noise corrections are applied. 
A 16 bin pattern artefact originating from the digitizer is removed from single-file $G_{12}^{(2)}$ traces prior to summation. 
A digital low-pass filter with a cutoff at 200\,MHz is applied to suppress high-frequency noise. 
Night-specific spectral noise profiles, obtained by FFT analysis of normalised correlations, reveal prominent interference from radio and television broadcasts (see \citenum{rockelein_analyse_2025}). 
Identified peaks with excess noise are attenuated using notch filters.
Specifically in the data for the night of July 18th, a large number of files are excluded due to clouds covering the star. A large increase in noise in the correlation files is visible, strongly deviating from the (otherwise very stable) expected noise behavior primarily depending on zenith angle.
To check the noise behavior after applied corrections, the cumulative $g^{(2)}(\tau)$ rms of the full dataset of all six nights is investigated excluding the regions in the $g^{(2)}(\tau = 0)$ containing the bunching peak and correlated electronic noise at zero delay. Figure \ref{fig:cumrmsplot} shows this noise behavior agreeing with the expected $1/\sqrt{N}$ shape of incoherent photons.

\begin{figure}[!h]
    \centering
    \includegraphics[width=0.9\columnwidth]{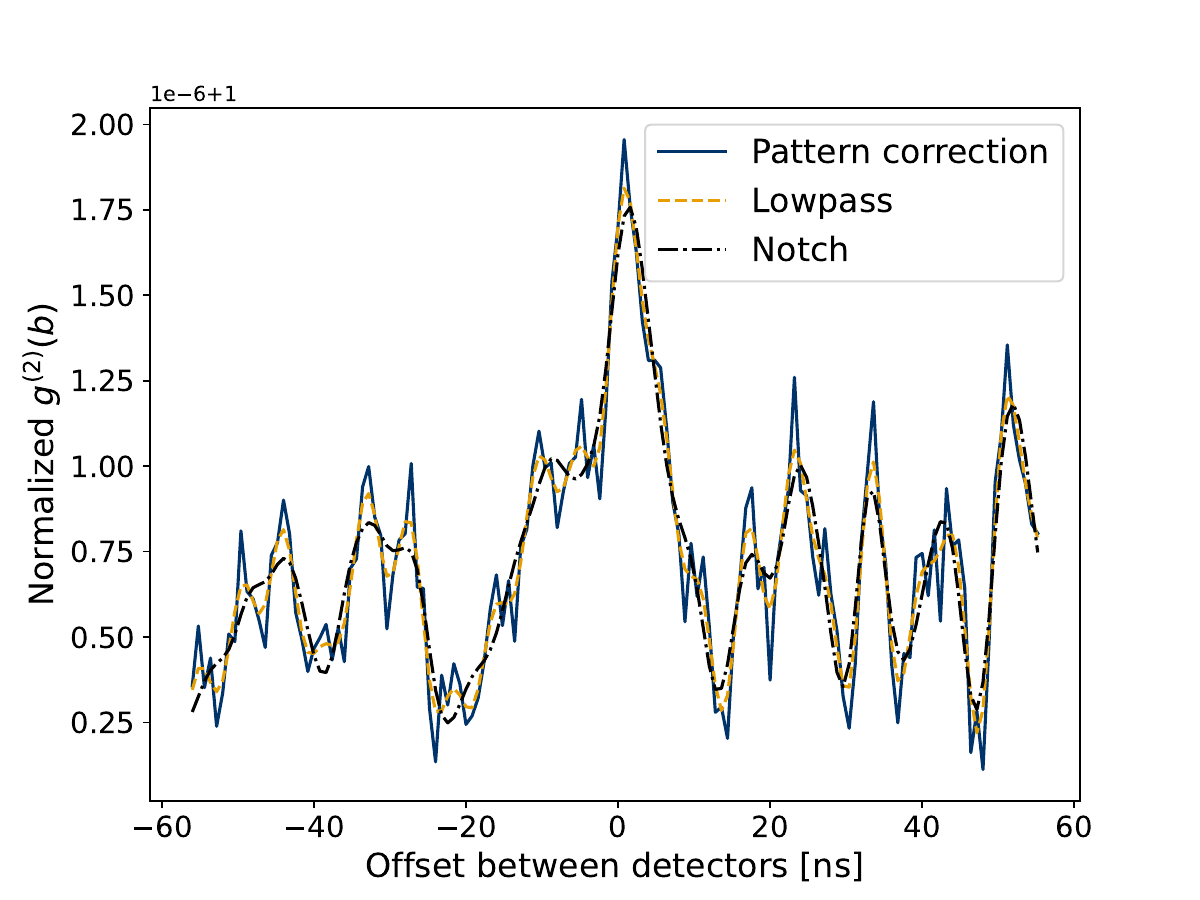}
    \caption{Evolution of synthetic peak region at different stages in noise correction. Blue shows the initial state of the data after pattern correction, orange after digital low-pass filtering, and black after applying notch filters (at frequencies specific to each respective night).}
    \label{fig:coreffects}
\end{figure}

To quantify the impact of these corrections on photon bunching, a synthetic template is constructed from Gaussian-smeared single-photon pulse shapes, calibrated nightly. 
This template, inserted into a segment of a representative $g^{(2)}(b)$ function, enables simulation of the effects of each correction step. 
This insertion is done far from zero delay to avoid the strongly correlated internal electronic noise of the digitizer card. 
As shown in Figure \ref{fig:coreffects}, while notch filtering produces minimal attenuation (less than $3\,$\% peak integral loss), the low-pass filter has the most pronounced smoothing effect on the bunching signature.

Two different methods are used to extract the photon bunching peaks from the summed and normalized $g^{(2)}(b)$ functions of each night or combination of nights:
A four-parameter Gaussian fit (or unfixed-sigma Gaussian fit as described in \citenum{ingenhutt_intensity_2025} section 4.1.1) and a template fit using correlated calibration-derived peak shapes (calibration peakshape fit in \citenum{ingenhutt_intensity_2025} section 4.1.3) are applied. 

Using the initial four-parameter Gaussian fit, attempts are made to fix three of the parameters (peak width, time offset, and correlation offset), and their effect on signal-to-noise ratio (SNR) is explored.

Fixing the peak width (sigma), as done in high-statistics experiments such as H.E.S.S., is not feasible for the MI\textsuperscript{2}SO dataset due to the large uncertainty in the resulting sigma obtained by this method. 
Moreover, potential changes in the instrument configuration across observation campaigns limit the assumption of constant time resolution throughout the data set. 
Particularly the two-month gap between the first observation nights and the September dataset calls assumptions of constant time resolution (or other factors) into question as the telescopes along with all measurement electronics had to be fully dismounted and removed from the institute roof for storage during bad weather. 
Although not significant, there are indications that the time resolution is not consistent between the July and September measurements. We decide not to fix the peak width for this reason as well as the high relative uncertainty resulting from the low statistics.

Attempts to fix the bunching peak position to the theoretically expected inter-detector delay calculated from cable lengths and signal propagation speed are impeded by residual jitter, even in the case of oversampled optical path delay correction. 
Consequently, this parameter is left unfixed in all fits, but constrained to a $4\,$ns window.

Efforts to fix the bunching-peak offset to unity, as predicted by the Siegert relation, are also explored. 
However, low-frequency noise sources, notably in the kHz regime from a nearby illuminated art installation, introduce distortions visible as a bend across the entire $40\,\upmu$s length of the correlation. 
Despite corrective filtering, night-to-night variations in the fitted offset remain comparable in magnitude to the amplitude of the bunching signal itself. 

As none of the attempted parameter fixing yielded improved SNR, all of them were left free for analysis via Gaussian fitting.

\begin{figure}[!h]
    \centering
    \includegraphics[width=0.9\columnwidth]{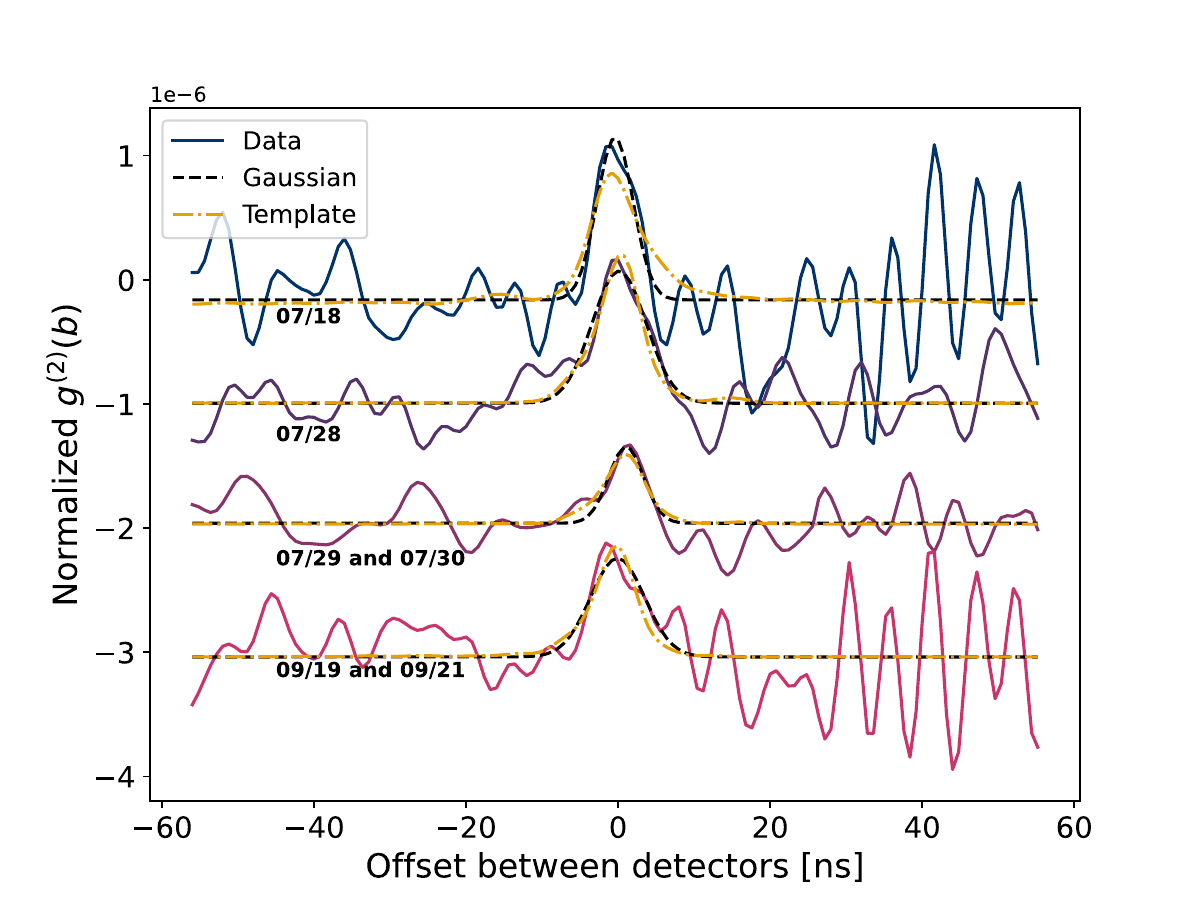}
    \caption{Comparison of photon bunching peak extraction via four-parameter Gaussian fit (black) and calibration peakshape template fit (orange) on all datasets' normalized summed $g^{(2)}(b)$. The four different datasets were plotted with an arbitrary offset to separate them in the figure.}
    \label{fig:gausspscomp}
\end{figure}

For the template fit, expected bunching peak shape templates are constructed by extracting the average single-photon pulse shapes in each detector from the nightly recorded calibration files, smearing these pulse shapes with a Gaussian using the measured detector TTS (see \citenum{blass_charakterisierung_2024} for further details) and finally correlating the two smeared single-photon pulse shapes. The resulting templates closely match the shape of measured photon bunching peaks and yield improved SNR over the four-parameter Gaussian model.
Figure \ref{fig:gausspscomp} shows a comparison between the two applied methods applied to the normalized summed $g^{(2)}(b)$ of each of the four datasets. At detector offsets (corrected for the cable length difference) greater than $\sim\!20\,$ns, one can see the increasingly large amplitudes of correlated noise due to crosstalk. Considering this strong effect, a larger difference in cable length and a switch to a digitizer card with less crosstalk should be considered for future measurements of this type.

\subsection{Results}\label{sec:results}
\begin{figure}[!h]
    \centering
    \includegraphics[width=0.9\columnwidth]{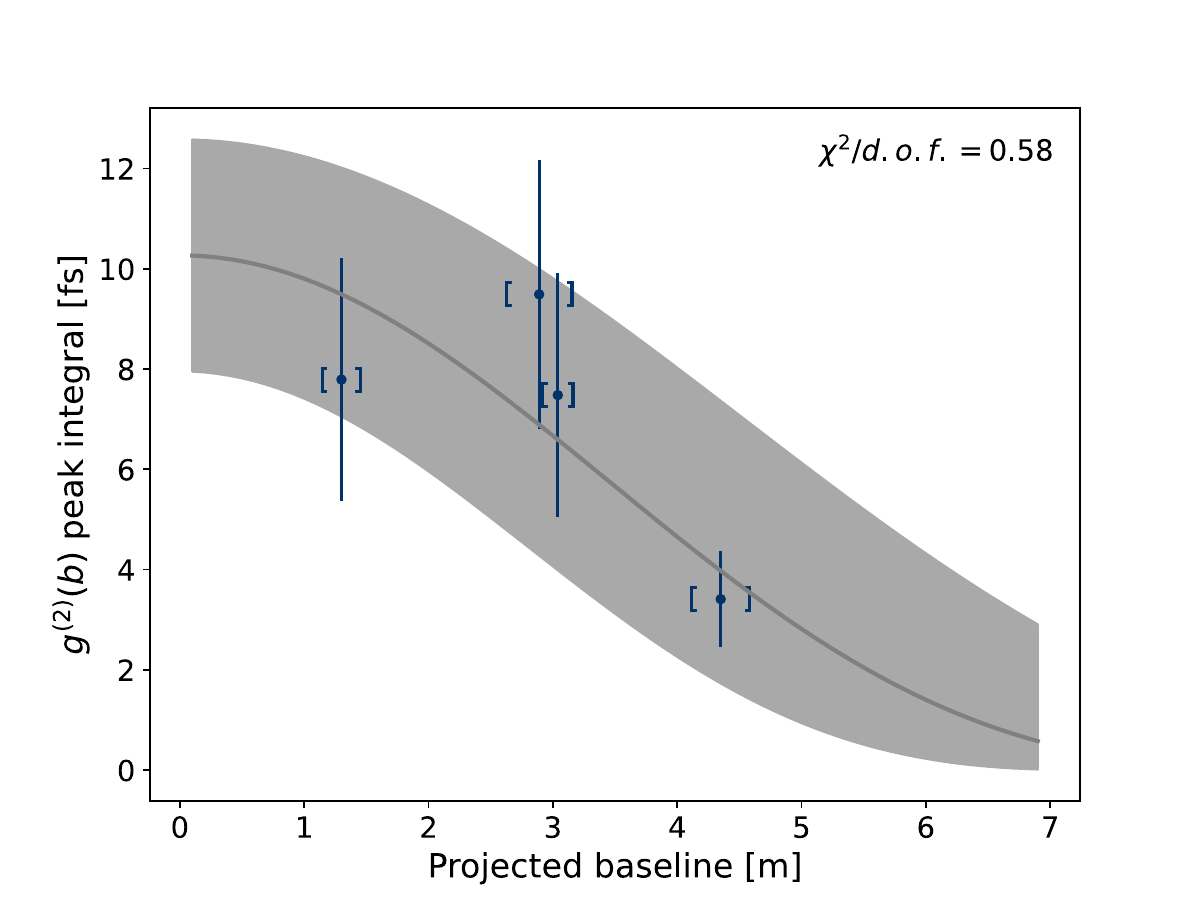}
    \caption{Spatial coherence curve fitted to photon bunching peak integrals extracted from a four-parameter Gaussian fit to the bunching peak. Brackets denote the range of baselines of the telescopes at the recording time of each measurement file integrated for each respective data point.}
    \label{fig:4paramsc}
\end{figure}

Using the four-parameter Gaussian fit, the angular diameter of Arcturus was determined to be $18.5 \pm 4.3$ milliarcseconds (mas) at $\lambda_0 = 655\,$nm. Figure \ref{fig:4paramsc} shows the spatial coherence fit. This fit was performed using only uncertainties in spatial coherence from the peak extraction. Square brackets denote the range of projected baselines resulting from the star's path in the sky relative to the (during each night) stationary telescope positions and are not statistical uncertainties used for the fit. Only uncertainties in spatial coherence resulting from the peak extraction via Gaussian fitting were used.

\begin{figure}[!h]
    \centering
    \includegraphics[width=0.9\columnwidth]{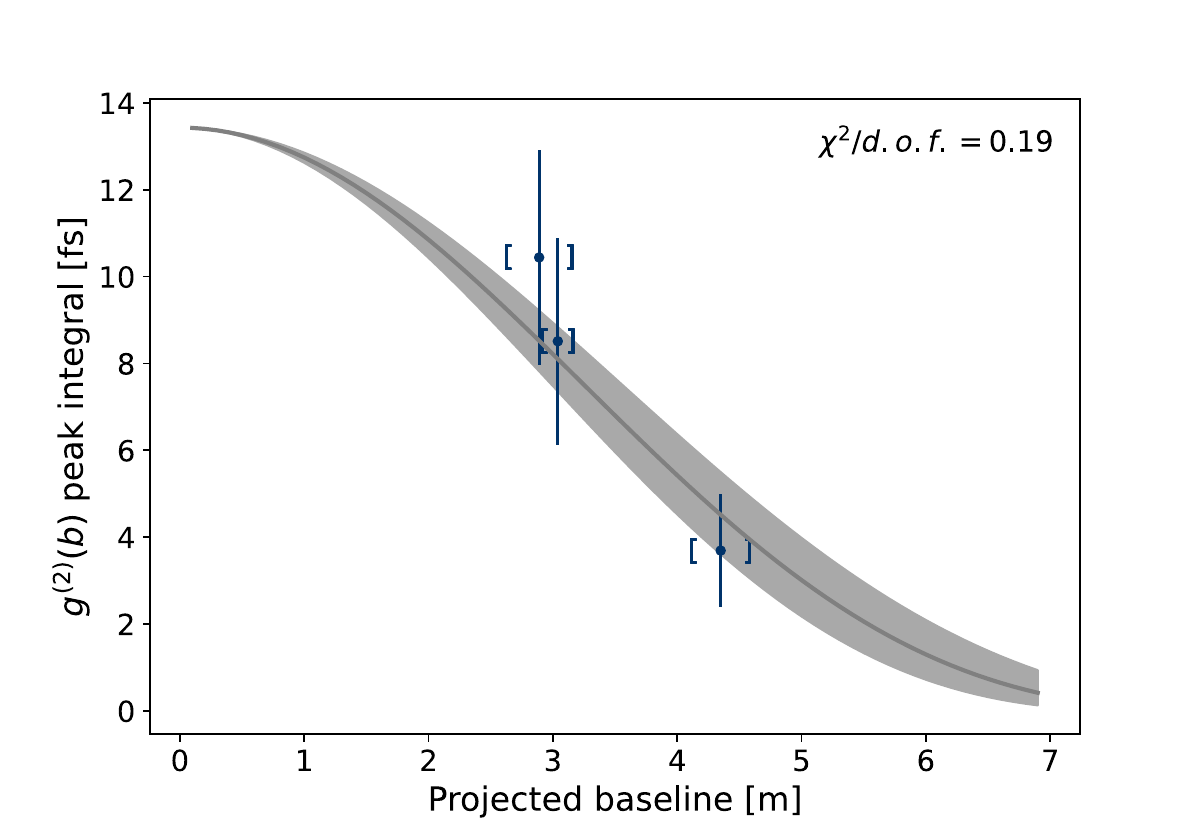}
    \caption{Spatial coherence curve fitted to photon bunching peak integrals extracted from a calibration peakshape fit to the bunching peak. Brackets denote the range of baselines of the telescopes at the recording time of each measurement file integrated for each respective data point. For illustrative purposes, September measurements were excluded from this analysis and spatial coherence at zero baseline was fixed to the theoretical coherence time calculated using the instrument's characteristics.}
    \label{fig:pssc}
\end{figure}

In Section \ref{sec:analysis}, concerns were outlined regarding the consistency in performance (most notably time resolution) of the setup between the July and September measurements. For illustrative purposes, the peaks extracted via the calibration peakshape method were analyzed excluding the September data in addition to the normal analysis using all data. Furthermore, the coherence time at a projected baseline of 0 can be theoretically calculated using the combined star and filter spectrum and geometry of the system. This theoretical coherence time was used as a fixed point for the fit. This analysis is shown in figure \ref{fig:pssc}.

Both figures include $1\,\sigma$ confidence regions around the fit as well as measured coherence time errors. The result of $19.7 \pm 1.7$ mas for the angular diameter of Arcturus is consistent with the four-parameter Gaussian fit and shows a smaller uncertainty resulting from the parameter fixing, data exclusion and change in peak extraction method. Both measurements agree within $1\,\sigma$ of the literature value of 19.77\,mas at 700\,nm\cite{ramirez_fundamental_2011}. 

Due to the low total amount of data recorded on Vega as well as the orientation of the telescopes relative to each other leading to projected baselines with very similar expected $g^{(2)}$ measurements, no further analysis was performed on the recorded data for this star. 

\section{Conclusions}\label{sec:conclusion}
We designed and built two optical Fresnel lens telescopes to use as an intensity interferometer. Focussing on keeping the instruments lightweight, the MI\textsuperscript{2}SO telescopes demonstrate mobility, allowing for changes in baseline and orientation on short timescales. Measurements of a bright star (Arcturus) yielded results for its angular diameter, in agreement with the values of the literature at milliarcsecond precision in $\sim\!10\,$h of total observation time. Instruments capable of such high angular resolution astronomy at optical wavelengths are (often due to their construction and operation costs) very limited in the availability of observation time for targets of opportunity. Some limitations were encountered during the measurement campaign, primarily radio and TV broadcast frequencies producing electronic noise within the expected signal frequency bands, as well as particularly difficult to track down systematics caused by repeated taking down and setting up of the telescopes due to weather. 

\begin{figure}[!h]
    \centering
    \includegraphics[width=0.9\columnwidth]{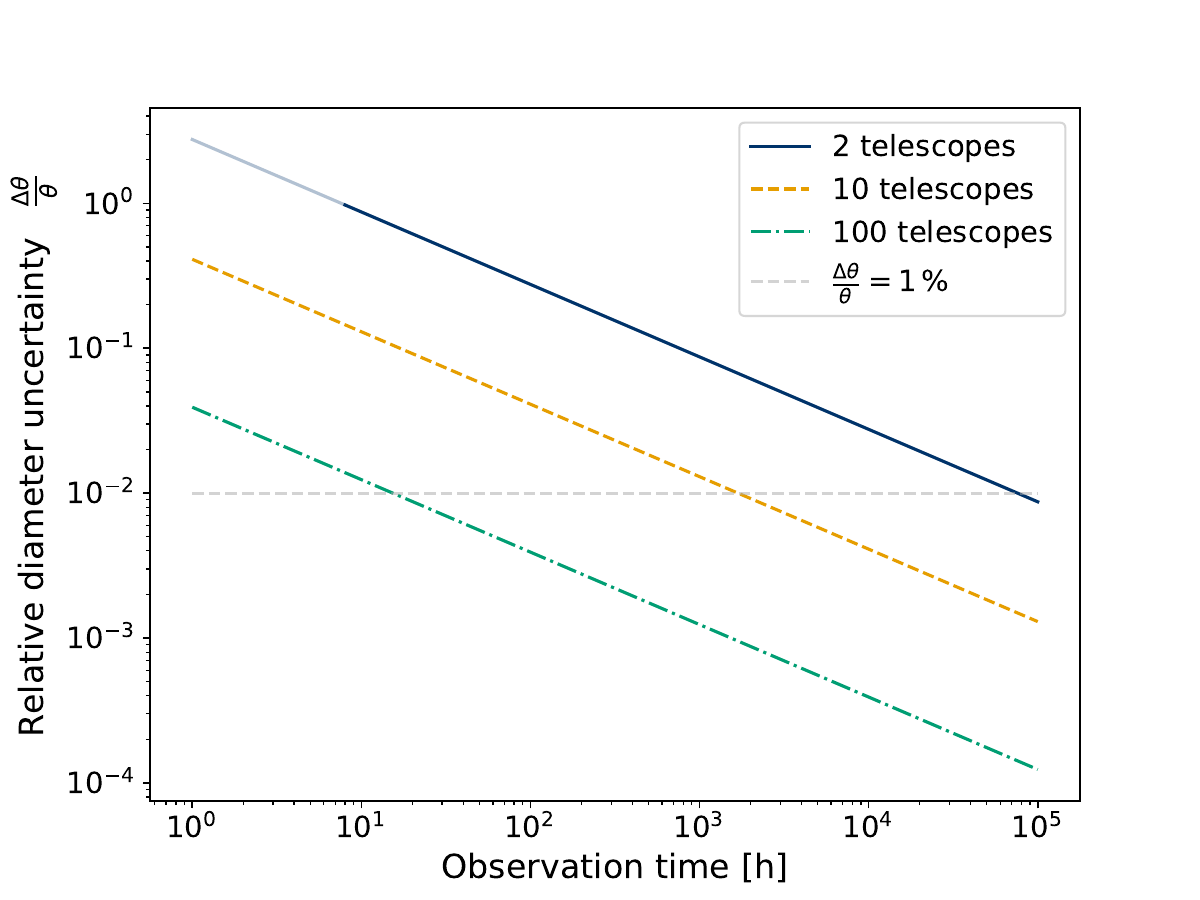}
    \caption{Estimated uncertainties on a diameter measured for a magnitude 2 star in a given observation time for 2, 10 and 100 MI\textsuperscript{2}SO telescopes. At this limiting magnitude, it would take a few nights of observations ($\sim 20\,$h of observation time) to determine a star's diameter to $1\,\%$ precision with $100$ telescopes.}
    \label{fig:sigprojection}
\end{figure}

Nevertheless, the MI\textsuperscript{2}SO demonstrated that low complexity, cheap ($\sim\!25000\,$\EURdig~for one unit) and lightweight ($\sim\!11\,$kg) telescopes using Fresnel lenses are potential instruments for SII. Figure \ref{fig:sigprojection} shows a projection of the uncertainty on a diameter measured for a magnitude 2 star. This was calculated for 2, 10 and 100 MI\textsuperscript{2}SO telescopes to illustrate the effect of quantity scaling relative to using more observation time. Uncertainties larger than $100\,\%$ were plotted with reduced alpha to illustrate that at these observation times, 2 telescopes would not be able to make any claims about the diameter of the star observed. With 100 telescopes, $1\,\%$ precision could be achieved on magnitude 2 stars within $\sim\!20\,$h of observation time. There are other factors that would also significantly improve the capabilities relative to this first version of MI\textsuperscript{2}SO. In particular, advances in detector technology making sub-nanosecond time resolution more affordable and processing speed for live correlations removing the need for large data storage capacity could significantly reduce the already low cost, as well as improve the performance of the telescopes in the future.

Although such small collection area telescopes will undoubtedly remain limited in sensitivity for faint stars---observations of stars above magnitude $\sim\!2$ are unrealistic without a significant improvement in detector time resolution---baseline variations of tens of centimeters over the course of a night were already achieved on static mounts. With the ability to measure at such precise, arbitrary, and stable baselines with a small baseline uncertainty resulting from the smaller collection area, instruments like MI\textsuperscript{2}SO should be suited to selectively investigate specific parts of the projected viewing plane (or u-v plane) of a target, providing a unique look into features in the spatial coherence unresolvable by static, large collection area telescopes.

It should be noted that the absolute resolution of an Intensity Interferometer does not depend on the number of telescopes nor the observation time, the limiting factor is the distance at which one can synchronize digitizers or time-to-digital converters and correlate their data streams. At present, synchronization to $\sim\!50\,$ps at distances of up to $10\,$km is achievable with fiber-connected White Rabbit nodes. At optical wavelengths, projected baselines of this order would allow for resolutions down to tens of $\upmu$as. Because targets bright enough to be measured in small $1\,$m diameter telescopes are not expected to be tens of $\upmu$as in angular diameter, it is best to think about intensity interferometers with mobile telescopes as having a somewhat arbitrary resolution range and instead a limiting magnitude above which the signal becomes too weak. 

For the targets that are bright enough for mobile telescopes, the dense coverage and arbitrary choice of projected baselines allows for detailed investigations of for example spherical asymmetries of a rapid rotator, orbital parameters of a binary system or stellar accretion. The alternative of using observation time on larger, stationary telescopes whose primary purpose is not SII would instead depend on serendipitously reaching the required projected baselines at the timing for which observation was granted. Therefore, dedicated intensity interferometers with mobile telescopes present a unique opportunity to test models of both single and multiple star system evolution.

\section*{Disclosures}
The authors declare that they have no financial interests, commercial affiliations, or other potential conflicts of interest that could have influenced the objectivity of this research or the writing of this paper.
\section*{Acknowledgements}
We thank William Guerin for his input on data analysis and uncertainty estimations in II.We acknowledge the financial and infrastructural support of the Erlangen Centre for Astroparticle Physics making this project possible.

\section*{Code and Data Availability}
The code and data underlying this article will be shared on reasonable request to the corresponding author. Correlation histograms are available in time intervals of $1.718\,$ s. Due to the large size of the digitized waveforms in excess of 100 TB, the raw data cannot be made available online.



\bibliographystyle{IEEEtran}
\bibliography{references} %





\vspace{2ex}\noindent\textbf{Christopher Ingenh\"utt} is a PhD student at the Astro Quantum Optics group at the Erlangen Centre for Astroparticle Physics, Friedrich-Alexander-Universit\"at Erlangen-N\"urnberg. He received his BS degree in physics in 2021, his MS degree with physics, specializing in astrophysics and astroparticle physics in 2025. His research interests include stellar intensity interferometry, instrument design, and binary system modelling.

\vspace{2ex}\noindent\textbf{Pedro Batista} is a postdoctoral researcher at the Erlangen Centre for Astroparticle Physics, Friedrich-Alexander-Universit\"at Erlangen-N\"urnberg. He has received his BS and MS degrees in physics from the University of São Paulo (USP), at the Institute of Physics of São Carlos (IFSC), in 2016 and 2019 respectively. He did his Ph.D. in the Gamma-ray group of the Deutsches Elektronen-Synchrotron (DESY) institute, Zeuthen Campus, while enroled with the University of Potsdam (UP).

\vspace{1ex}
\noindent Biographies and photographs of the other authors are not available.
\label{lastpage}
\end{document}